\documentclass{article}
\usepackage{graphicx}
\usepackage{epstopdf}
\usepackage[all]{xy}
\usepackage{amsmath, amssymb,amsfonts,amsthm,amsbsy}
\newtheorem{theorem}{Theorem}[section]

\newtheorem{proposition}[theorem]{Proposition}

\newenvironment{definition}[1][Definition]{\begin{trivlist}
\item[\hskip \labelsep  {\bfseries #1} ]}{\end{trivlist}}

\newcommand{\bc}{\begin{center}}
\newcommand{\ec}{\end{center}}
\newcommand{\be}{\begin{equation}}
\newcommand{\ee}{\end{equation}}
\newcommand{\bea}{\begin{eqnarray}}
\newcommand{\eea}{\end{eqnarray}}
\newcommand{\beq}{\begin{eqnarray*}}
\newcommand{\eeq}{\end{eqnarray*}}

\begin{document}


\title{Factorising  equity returns in an emerging market through exogenous shocks and capital flows}

\author{Diane Wilcox\footnote{School of Computational and
Applied Mathematics, University of the Witwatersrand,
South Africa, {\tt diane.wilcox@wits.ac.za} (corresponding author) } \ and \ Tim Gebbie }


\date{22 June 2013}

\maketitle

\begin{abstract}
A technique from stochastic portfolio theory [Fernholz, 1998] is applied to analyse equity returns of Small, Mid and Large cap portfolios
in an emerging market through periods of growth and regional
crises, up to the onset of the global financial crisis. In
particular, we factorize portfolios in the South African market in
terms of distribution of capital, change of stock ranks in
portfolios, and the effect due to dividends for the period Nov
1994 to May 2007. We discuss the results in the context of broader
economic thinking to consider capital flows as risk factors, turning around more established approaches which use macroeconomic and socio-economic conditions
 to explain Foreign Direct Investment (into the economy) and Net Portfolio Investment (into equity and bond markets).

 \end{abstract}


%
\begin{center}{\footnotesize{Keywords: stochastic portfolio theory, risk factors, foreign direct investment, net portfolio investment, size-effect, emerging markets, fragility}}
\end{center}


\section{Introduction}

Risk factors driving equity portfolio returns may include
performance measurements  relative to some market index as in the
capital asset pricing model (CAPM \cite{Lintner, Sharp, Mossin,
Treynor} and its offspring) or more general arbitrage pricing
theory (APT, \cite{Ross}) risk factors in neo-classical finance.
Some portfolio managers may select securities based on performance
variables such as earnings-per-share or price-to-book estimates
while others may be interested in valuations which are explained
in terms of macroeconomic factors such as interest rates,
inflation, GDP, market index or forex levels. The \emph{size
effect} refers to empirical evidence that small-capitalised stocks
exhibit higher long-term performance than larger capitalised
stocks
 \cite{Banz}.
This led to the development the  Fama and French model to  explain
market, size and value (\cite{Ball, Basu}) components of equity
returns by means of three APT-consistent risk factors
\cite{FF1992}.

\vspace{0.3cm}
Historically a simple motivation for the size effect is that some
concerns with small capitalisation have potential for growth which
may be far greater than any large cap stock. According to
contemporary  classifications in the US market, stocks may be
classified in the following size-based scheme:

\vspace{0.3cm}
\begin{tabular}{|c|c|}
    \hline
Mega cap & 200 billion USD and greater \\
  \hline
Big cap  & 10 billion  USD and greater \\
 \hline
Mid cap & 2-10 billion USD \\
\hline
Small cap & 300 million - 2 billion USD \\
\hline
Micro cap & 50-300 million USD \\
\hline
Nano cap & under 50 million USD \\
\hline
\end{tabular}

\vspace{0.3cm}According to the global classification, in 2013 the
top  14 stock on the JSE are Big caps with the next 2 on the
Big-Mid boundary. The largest 70 JSE stocks are global Mid caps,
while the remaining top 100 are all worth more than 1 billion USD,
placing them within global Small cap range.

\begin{figure}
\centering
\includegraphics[width=10cm]{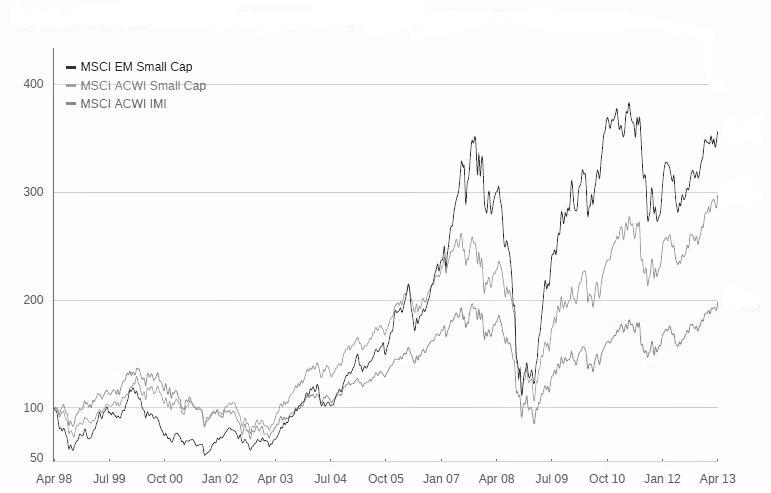}
\caption{At the peak of equity markets in 2007, the MSCI index
for Emerging Market (EM)  Small Cap outperforms the All Country
World Index (ACWI)  Small Cap which in turn outperforms the ACWI
Investable Market Index (IMI) before all collapse  to '98/'99
levels in 2008. The graphs depict cumulative index performance based on net returns.}\label{msci}
\end{figure}

\vspace{0.3cm} Capitalisation of stocks in the SA equity market
peaked at 833,548 million USD in 2007, dropped in 2008 and rose to
1,012,540 million USD by  2010 before dropping back to 2007 levels
in 2011. This volatile behaviour has obviously been driven by the
global financial crisis (GFC). Small cap indices show
exacerbated fragility through the GFC, where values more than
tripled before collapsing to 1998 levels by the end of 2008.
Figure ~\ref{msci} plots the course of the MSCI Emerging Markets
Small Cap Index, ``which includes small cap representation across
21 Emerging Markets countries"\footnote{including  Brazil, Chile,
China, Colombia, Czech Republic, Egypt, Hungary, India, Indonesia,
Korea, Malaysia, Mexico, Morocco, Peru, Philippines, Poland,
Russia, South Africa, Taiwan, Thailand and Turkey}, with 1,775
constituents. This index ``tracks roughly 14\% of the (free
float-adjusted) market capitalization in each country."  Clearly
oscillations in cross-border portfolio flows have nontrivial
effects on conditions in  a developing market (see also \cite{McC2012}) and a better
understanding of small cap stocks is relevant for investments of
various horizons particularly during periods of quantitative easing and withdrawal thereof \cite{Eich}.

\vspace{0.3cm} Underlying reasons for the size-effect continue to
be  debateable. One plausible explanation  came from the so-called
stochastic portfolio theory (SPT) of Fernholz \cite{FS1982}. In \cite{GW2008}
we reviewed the SPT rank-based construction and showed
outperformance of small cap  portfolios can be obtained in purely
Gaussian price formation models. Under a simplifying assumption
(discussed in Section ~\ref{sizeprf}), it is easy to generate a 
 nullcase in which the selection mechanism, and not some
 meaningful performance related attribute, can explain
outperformance in a random price market. This suggests that 
arguments which attribute size to being a proxy for some risk
premium such as liquidity or capacity for firm-related growth in
the economy are only part of the story.

\vspace{0.3cm} When the model was proposed for US equity markets
in 1999, Fernholz was interested in isolating a size factor\cite{F1998,F2001}. In
that research, the effect was isolated in a market which, by 1999,
had seen significant capital growth just before the violent swings
since the dotcom bubble.  In \cite{F2002}  60yrs of data,
1939-1999, are examined to isolate a size effect. Fernholz's
empirical analysis of so-called {\em distributional}  components
covered the period 1989-1999  \cite{F2002}. The period
1989-1999 is also used to examine components of returns in a
portfolio generated via  ranking in value. In that analysis, the
distributional component is noted to be  correlated to an
entropy-based measure of diversity. The findings were similar in
the paper \cite{F1998} for the period 1975-1999.

\vspace{0.3cm} We claim that  SPT is particularly useful as a  lens in emerging markets  to separate
market capital distribution from investor preferences based on
ranking. In particular,  the distributional, rank and dividend
components of a SPT factorisation offer useful maps to capital
flow, size (by market cap or other rank-based investment criteria)
and profit taking factors to explain equity prices. This suggests
 a novel perspective toward unravelling the nature of portfolio
flows as a risk factor in financial markets.

 \vspace{0.3cm}
While there are investigations to uncover the economic  factors
which drive cross-border cashflows \cite{RR2003} complemented by critical discussion thereof \cite{Ashman, Mohamed}, to our knowledge
there is very little empirical work which considers the effect of
net portfolio investment (NPI)  as a macroeconomic factor driving growth in a
market sector. In \cite{JF2011}, the relationship between weekly equity returns for JSE listed stock between Jan 2002 and Dec 2006 and NPI is investigated. A  (linear) vector autoregressive model is used to identify lead and lag effects between these factors. In that work evidence is found that returns on the JSE tend to forecast NPI and find an absence of an impact of NPI on returns.

\vspace{0.3cm} In this paper we document the SPT decomposition of
portfolios constructed via size and value criteria. We also
observe that significant changes in components are indicative of
changes of states which, as expected, appear to be correlated to
changes in levels of cross-border portfolios flows and investment.
 In our investigation we focus on price formation in the South African (SA)
equity market for the  period Nov 1994 to May 2007. In this
post-democracy to pre-GFC period, SA market pricing has been
impacted by numerous exogenous events including three ZAR currency
crises, the Asian crisis of 1997, the dotcom bubble, the USA 9/11
Twin Trade tower collapses in 2001, the 2nd Iraq war and the
resources and property booms up to 2007. Taking into account tacit
market knowledge about features such as liquidity and bid-offer
spreads for specific stocks between 1994 and 2007, we construct
Small, Mid and Large cap portfolios by setting thresholds in the ranked
market caps. We also use book-value-to-price information to
construct a Value portfolio for complementary analysis. Our investigation exposes correlations between regime changes in investment trends in JSE equities and NPI. This is in keeping with post GFC developments to better address the failures of orthodox economic theory and to better account for imperfect knowledge \cite{Lux, SvH}.

\section{Continuous time stochastic rank-based \\ capital distribution}

We recap some notation and key results  for describing the main
decomposition \cite{F2001, F2002}.

\subsection{Growth rates and rates of return}

The continuous time formulation of the SPT framework  makes it
compatible with other continuous time models used in pricing
securities. Generalisation of the simpler discrete time analysis
depends on applications of It$\hat{o}$'s lemma and a
representation of rank crossovers in terms Tanaka's formula for
Brownian local times.

\vspace{0.3cm}The use of market capitalisation (cap) instead of
price in the SPT modelling framework facilitates measurement of
capital distribution through the market. For the purpose of this
discussion we assume a market without corporate events such as
mergers and acquisitions, spinoffs, etc\footnote{This approach
also dispenses with the need to adjust for stock splits and
historic data can be filtered for such cases.}. The construction
of size-based portfolios then becomes a easy exercise in ranking
stocks from biggest to smallest cap and determining a threshold to
separate  large from small. The continuous time formulation is
able to quantify activity at the boundary through (stochastic)
{\em local time} measurement of multivariate processes.

\vspace{0.3cm}

A stock is modelled by its market cap, where  the  total value of
shares issued by the $i^{th}$ corporation is denoted $X_i$. We
sometimes refer to this as the stock price, a shareholder may hold
a fraction of this share and prices are assumed to be driven by an
idiosyncratic drift coupled to a market multivariate Wiener noise
process:
 \cite{FS1982}:
\begin{eqnarray}
 d\log X_i(t) = & \underbrace{\gamma_i(t) dt}& + \sum\limits_{\nu=1}^{n}\xi_{i\nu}(t)dW_{\nu}(t) \hspace{3.5cm}\\
 &\textrm{growth rate}&  \nonumber
\end{eqnarray}
where  $(W_1, ..., W_n)$ is standard Brownian motion, $\gamma_i$
is measurable, adapted and for all $T>0$, $i=1,...n$, satisfies
the condition $\int_0^T|\gamma_i(t)|dt<\infty$ a.s.,  the
$\xi_{i\nu}$, $\nu=1,...n$ are measurable, adapted and satisfy
$\int_0^T \sum\limits_{\nu=1}^{n}\xi_{i\nu}^2(t) dt<\infty$ for
all $T>0$, a.s. and
$\lim\limits_{t\rightarrow\infty}\frac{1}{t}\sum\limits_{\nu=1}^{n}\xi_{i\nu}^2(t)\log\log
t = 0$. The processes $\gamma_i$  and $\xi_{i\nu}$ represent the
{\it growth rate} and the volatility of the $i^{th}$ stock,
respectively.

An application of  It$\hat{o}$'s lemma leads to
\begin{eqnarray}
d X_i(t) =  &\underbrace{\left(\gamma_i(t)+\frac{1}{2}\sum\limits_{\nu=1}^{n}\xi_{i\nu}^2(t)\right)}&X_i(t)dt + X_i(t)
\sum\limits_{\nu=1}^{n}\xi_{i\nu}(t)dW_{\nu}(t), \\
&\textrm{rate of return}& \nonumber
\end{eqnarray}
where $\alpha(t) = \gamma_i(t) + \frac{1}{2}\sum\limits_{\nu=1}^{n}\xi_{i\nu}^2(t)$ is referred to as the \emph{rate of return} process.

According to Fernholz,  while the rate of return is used in
classical portfolio theory, the growth rate is a better indicator
of long-term behaviour.  As a special case, it is notable that if
growth rates are assumed to be constant, then long-term
equilibrium  will exist only if growth rates are the same for all
stocks in the market. If any stock were to have a higher growth
rate relative to the rest, then that stock would eventually
dominate the market and stocks with lower growth rates would
disappear. In reality, growth rates are not constant and markets
are more likely to be impacted by global events outside the model.

\subsection{Portfolios and portfolio growth rates}

 The {\it market} is viewed as the family of all stocks
 $X_1,\ldots,X_n$, where two stocks are equivalent if they differ
by at most a constant. A {\it portfolio} $\pi$ is defined to be a
positive-valued combination of stocks such that
\begin{eqnarray}
Z_{\pi} = \pi_1 Z_{\pi} + \ldots \pi_n Z_{\pi},
\end{eqnarray}
where the $\pi_i$ are functions of time and denote the proportion
of $Z_{\pi}$ invested in $X_i$. The quantity $\pi_i Z_{\pi} $ is
the amount invested in $X_i$ and thus, a portfolio $\pi$ can
represented by the weights allocated to each stock, $\pi_1,
\ldots, \pi_n$, at time $t$. It is assumed that these weights are
bounded and sum to unity, $\sum \pi_i = 1$, so that the portfolio
is fully invested in stocks.

It is also useful to differentiate between a {\it passive
portfolio}, where the fractional number of shares held in each
stock are constants: $ Z_{\nu} = \nu_1 X_1 + \ldots \nu_2 X_2$,
and a {\it balanced portfolio} where the proportions $\pi_i $ are
constant. The proportions can be related to the fractional number of units held
as $\pi_i = \nu_i X_i / Z_{\nu}$. Here constant values for
$\nu_i's$ imply that the proportions $\pi_i$ are not necessarily
constant.


\begin{definition}
\emph{(Market Portfolio)} The market portfolio $\mu$ has weights $\mu_1, \ldots, \mu_n$
given by:
\begin{eqnarray}
\mu_i(t) = \frac{X_i(t)}{X_1(t)+\ldots+X_n(t)}, \label{eqn:marketport}
\end{eqnarray}
for  $t \in [0,T]$, $i=1,\ldots,n$;  the
weights $\mu_i(t)$ are called the market weights.
\end{definition}

 The value of the market portfolio,
\begin{eqnarray}
Z_{\mu}(t) = X_1(t) + \ldots + X_n(t),
\end{eqnarray}
 is the combined capitalisation of all the
stocks in the market at a given time.

From the definitions of the price processes of stocks, it follows that
\begin{eqnarray}
 d\log Z_{\pi}(t) =  &\gamma_{\pi}(t)dt + \sum\limits_{i,\nu}\pi_i(t)\xi_{i\nu}dW_{\nu}(t), \hspace{4.5cm}& \\
\textrm{ where} \hspace{0.5cm} \gamma_{\pi}(t) =  &\underbrace{\sum_i \pi_i(t)\gamma_i(t) + \frac{1}{2}\left( \sum_i \pi_i(t)\sigma_{ii}(t) - \sum_{i,j}\pi_i(t)\pi_j(t)\sigma_{ij}(t)\right)}& \nonumber \\
 &\textrm{portfolio growth rate}& \nonumber \\
 & & \nonumber \\
 \textrm{and} \hspace{0.5cm} \sigma_{ij}(t) = & \sum_{\nu}\xi_{i\nu}(t)\xi_{j\nu}(t)dt. \hspace{6.5cm} & \nonumber
  \end{eqnarray}

Equivalently,
\begin{eqnarray}
 d\log Z_{\pi}(t) &=& \sum_i \pi_i(t)\gamma_i(t)dt + \gamma_{\pi}^*(t)dt + \sum\limits_{i,\nu}\pi_i(t)\xi_{i\nu}dW_{\nu}(t), \\
\textrm{ where} \hspace{0.5cm}
\gamma^*_{\pi}(t) &=& \underbrace{\frac{1}{2}\left( \sum_i \pi_i(t)\sigma_{ii}(t) - \sum_{i,j}\pi_i(t)\pi_j(t)\sigma_{ij}(t)\right)}. \nonumber \\
& & \hspace{1.7cm} \textrm{excess growth rate}
\end{eqnarray}

The \emph{portfolio growth rate} exceeds the weighted average  of
the individual growth rates by the amount referred to as the
\emph{excess growth rate}. The logarithmic representation exposes
this insight. In the standard representation, the portfolio rate
of return is simply the portfolio weighted sum of the rates of
returns of the stocks comprising it.

\vspace{0.3cm}
Given a portfolio $\pi$, one may consider its performance relative
to the market:

\vspace{-0.2cm}
\begin{eqnarray}
d \log (Z_{\pi}(t)/Z_{\mu}(t)) = \sum_i \pi_i(t) d \log \mu_i(t) + \gamma_{\pi}^*(t)dt.
\end{eqnarray}

In particular, it is possible to understand the relative returns
 in terms of changes in the market weights and the excess growth rate process \cite{F2002}.

\subsection{Functionally generated portfolios}

Motivated by the relative performance of an entropy weighted
portfolio, Fernholz introduced the notion of a \emph{functionally
generated portfolio.} In this approach,  relative performance of a
portfolio can be decomposed into two distinct components within a
stochastic differential equation. The first component is the
logarithmic change in the value of the generating function, the
second term involves the relative covariances of the stocks in the
portfolio.

\vspace{0.2cm} A portfolio is constructed based on the entropy function:
\begin{eqnarray*} \mathbf{S}(x) = -\sum_{i=1}^{n}x_i\log x_i \end{eqnarray*} for all $x\in
\Delta^n = \{x\in \mathbf{R}^n | \sum x_i=1, x_i>0\}$. The
\emph{market entropy process} is defined by

\vspace{-0.3cm}
\begin{eqnarray}\mathbf{S}(\mu(t)) = -\sum_{i=1}^{n}\mu_i(t)\log \mu_i(t), \ \ t\in[0,T],
\end{eqnarray}
where $\mu$ denotes the market portfolio and the portfolio $\pi$
with weights
\begin{eqnarray} \pi_i(t) = \frac{\mu_i(t)\log \mu_i(t)}{\mathbf{S}(\mu(t))}, \ \ t\in[0,T], i=1...n,
\end{eqnarray}
is called the \emph{entropy weighted portfolio}.
 It can be shown that
 \begin{eqnarray}
 d \log \mathbf{S}(\mu(t)) = d \log\left(\frac{Z_{\pi}(t)}{Z_{\mu}(t)}\right) - \frac{\gamma^*_{\mu}(t)}{\mathbf{S}(\mu(t))}.
\end{eqnarray}

Generalising this construction is the idea of a \emph{functionally
generated portfolio}, which provides a decomposition of the
relative rate of return of portfolio weights into two distinct
components within a stochastic differential equation.

\begin{definition}
\emph{(Generating functions and functionally generated
portfolios)} Let $\mathbf{S}$ be a positive continuous function
defined on $\Delta^n$, and let $\pi$ be a portfolio.  $\mathbf{S}$
is said to generate $\pi$ if there exists a measurable process of
bounded variation $\Theta$ such that for all $t \in [0,T]$, a.s.

\begin{eqnarray}
 \log(Z_{\pi}/Z_{\mu}) = \log \mathbf{S} (\mu(t)) + \Theta(t), \label{eqn:relret}
\end{eqnarray}
The process $\Theta$ is called the
drift process corresponding to $\mathbf{S}$ If $\mathbf{S}$
generates $\pi$, then $\mathbf{S}$ is called the \emph{generating
function} of $\pi$, and $\pi$ is said to be \emph{functionally
generated.}
\end{definition}

 Equation (\ref{eqn:relret}) can be expressed in differential form as
\begin{eqnarray}
d \log( Z_{\pi}(t) / Z_{\mu}(t)) = d \log \mathrm{S}(\mu(t)) + d
\Theta(t), \label{eqn:312F2002}
\end{eqnarray}
a.s., for all $t \in [0,T]$. It is this differential form that is
typically used.

\subsection{Rank processes}

To investigate the size effect, one may consider a portfolio
generating function based on a market cap rank process.
Making the notion of rank mathematically precise, one may define the $k$th ranked number of the set $\{x_1, ...,x_n\} \subset \mathbb{R}$ to be
\begin{eqnarray} x_{(k)} = \max\limits_{1\leq i_1 < ...< i_k \leq n } \min (x_{i_1}, ...x_{i_k}), \ \ \ 1\leq k\leq n. \end{eqnarray}
It follows that
\begin{eqnarray}  x_{(1)} \geq x_{(2)} \geq ... \geq x_{(n)}. \end{eqnarray}

If $X_1, ..., X_n$ are stochastic processes, then the $k$th ranked process at time $ t\in [0,T]$ of
 $\{X_1, ....X_n\}$ is defined to be
\begin{eqnarray} X_{(k)}(t) = \max\limits_{1\leq i_1 < ...< i_k \leq n } \min (X_{i_1}(t), ...,X_{i_k}(t)), \ \ \ 1\leq k\leq n \end{eqnarray}

 To obtain mathematical expressions for the time-varying portfolio $\pi$ and drift process $\Theta$ for a generating
functions based on rank processes, one needs to appeal to \emph{local times} of
semi-martingales. These describe the occupation time that a process spends at a specific level. In the context at hand it is used to quantify transitions between large cap and small cap portfolios. Specifically, local times of rank differences at the boundary are instantaneously zero when there is a rank crossover.

\vspace{0.3cm}
For the continuous semi-martingale  $M$ in the next theorem, the local time process counts the number of times that $M$ hits zero..

\begin{theorem}
\emph{(Tanaka's formula for local times)} Let $M$ be a continuous semi-martingale. Then
the local time (at 0) for $M$ is the process $\Lambda_M$ defined
for $t \in [0,T]$ by
\begin{eqnarray}
\Lambda_{M(t)} = \frac{1}{2} \left( {|M(t)|-|M(0)| - \int_0^t
\textrm{sgn}(M(s)) dM(s)} \right), \label{tanaka}
\end{eqnarray}
where $\textrm{sgn}(m)= 2 I_{(0,\infty)}(m) -1$, with $I_{(0,\infty)}$
the indicator function of $(0,\infty)$.
\end{theorem}

\vspace{0.3cm}
Considering Equation (~\ref{tanaka}) in differential form and using the equivalences $\max(X,Y) = \frac{X+Y}{2} +\frac{|X-Y|}{2}$ and $\min(X,Y)=-\max(-X,-Y)$, it can be shown that rank processes derived from pathwise mutually
non-degenerate absolutely continuous semi-martingales can be
expressed in terms of the original process adjusted by local
times. As a corollary, one obtains the following result for market
weight processes:

\begin{proposition}
\label{prop:rankproc}Consider a market of stocks
$X_1,\ldots,X_n$ which are pathwise mutually non-degenerate. The
market weight processes $\mu_1, ... , \mu_n$, satisfy
\begin{eqnarray}
d \log \mu_{p_{t}(k)}(t) &&=  \sum_{i=1}^n I_{\{i\}}(p_{t}(k)) d \log \mu_i(t) \\
+&& \frac{1}{2} \left( {d\Lambda_{\log \mu_{p_{t}(k)}-\log
\mu_{p_{t}(k+1)}}(t)-d\Lambda_{\log \mu_{p_{t}(k-1)} - \log
\mu_{p_{t}(k)}}(t)}\right).
\end{eqnarray}
a.s, for $t \in [0,T]$, where $p_{t}$ is the random permutation
of $\{1,\ldots,n\}$, such that for $k=1,\ldots,n$.
\begin{eqnarray}
\mu_{p_t(k)}(t) &=& \mu_{(k)}(t), \\
 ~\mathrm{if}~ \mu_{p_{t}(k)}(t) &=&
\mu_{p_{t}(k+1)}(t),  ~\mathrm{then}~  p_{t}(k) < p_{t}(k+1)
\end{eqnarray}
\end{proposition}

\vspace{0.3cm}
Proposition ~\ref{prop:rankproc} is applied to obtain the
following key result which determines the portfolio weights $\pi$
and drift $\Theta$ for a portfolio generating function which
models an evolution of ranks through time.

\begin{theorem}
\emph{(Rank process generated portfolios)}
\label{thm:rankgenports} Let ${\cal M}$ be a market of stock
$X_1,\ldots,X_n$ that are pathwise mutually non-degenerate, let
$p_{t}$ be the random permutation defined as in Proposition \ref{prop:rankproc}.
 Let ${\mathbf
S}$ be a function defined on a neighborhood $U$ of $\Delta_n$.
Suppose that there exits a positive $S\in C^2(U)$ such that for $(x_1,\ldots,x_n) \in U$,
$\mathbf{S}(x_1,\ldots,x_n)= S(x_{p_t(1)},\ldots,x_{p_t(n)})$, and
for $i=1,\ldots,n$,  and $x_i D_i \log S(x)$ bounded for $x \in
\Delta^n$. Then $\mathbf{S}(x)$ generates the portfolio $\pi$ such
that for $k=1,\ldots,n$,
\begin{eqnarray}
\pi_{p_t(k)}(t) = \left( {D_k \log S(\mu_{p_t}(t))} + 1 - \sum_{j=1}^{n}{\mu_{p_t(j)} D_j \log S(\mu_{p_t}(t))} \right) \mu_{p_t(k)}(t),
\end{eqnarray}
for all $t \in [0,T]$, a.s., with a drift process $\Theta$.
\begin{eqnarray}
d \Theta(t) = \theta_{\pi}(t) dt + d L_{\pi}(t),
\end{eqnarray}
for all $ t \in [0,T]$, a.s. where, a.s.,
\begin{eqnarray}
\theta_{\pi}(t) = {-1 \over 2 \mathbf{S}(\mu(t))} \sum_{i,j=1}^n D_{ij} S(\mu_{p_t}(t)) \mu_{p_t(i)}(t) \mu_{p_t(j)}(t) \tau_{p_t(ij)}(t)dt
\end{eqnarray}
for $t\in[0,T]$,
 and
\begin{eqnarray}
 d L_{\pi}(t) = \frac{1}{2} \sum_{k=1}^{n-1} \left({\pi_{p_t(k+1)}(t) -\pi_{p_t(k)}(t)}\right)d \Lambda_{\log \mu_{p_t(k)} - \log \mu_{p_t(k+1)}}(t)
\end{eqnarray}
and for $t\in[0,T]$, $\tau_{ij}(t) = \sigma_{ij}(t) - \sigma_{i\mu}(t)- \sigma_{j\mu}(t) + \sigma_{\mu\mu}(t),$
$\sigma_{i\mu}(t) = \sum\limits_{j=1}^n\mu_j(t)\sigma_{ij}(t)$ and
$\sigma_{\mu\mu}(t)= \mu(t)\sigma(t)\mu^T(t)$.

\end{theorem}

\vspace{0.3cm}
The relative rank covariance process is denoted $\tau_{p_{t}(ij)} =
\tau_{p_{t}(i)p_{t}(j)}(t)$ for $i,j=1, \ldots n$ and the ranked
market weights $\mu_{p_{t}(1)},\ldots,\mu_{p_{t}(n)}$ generate
$\mu_{p_{t}} = (\mu_{p_t(1)},\ldots,\mu_{p_t(n)})$.
The first term of the drift, $\theta_{\pi}(t)dt$ is the
\emph{smooth} component and the second term, $dL_{\pi}$ is the
local time component.

\subsection{Size in continuous time}

We review the construction of portfolio generating
functions that partition the market  into mutually
exclusive portfolios of \emph{large capitalisation} and \emph{small
capitalisation}\footnote{We refer to a large (or small)
capitalisation portfolio simply as a large (or small) portfolio.} stock.

Let $ 1<m<n$ and consider the generating function for the large
portfolio:
\begin{eqnarray}
\mathbf{S}_L(x) = x_{p_t(1)} + \ldots + x_{p_t(m)}.
\end{eqnarray}
This represents the relative capitalisation of a large-stock index
composed of the $m$ largest stocks in the market. By
Theorem \ref{thm:rankgenports},  $\mathbf{S}_L$ generates a
portfolio $\xi$ with weights given by:
\begin{eqnarray}
 \xi_{p_t(k)}(t) = \left\{ {\begin{array}{ll} {\mu_{p_t(k)}(t) \over \mathbf{S}(\mu(t))},& k \leq m \\ 0, &k>m \end{array} } \right. \label{eqn:largewts}
 \end{eqnarray}
 for $t \in [0,T]$, where weights $\xi_i$ for $i=1,\ldots,m$ represent the capitalisation weights of the stocks in the
 large stock index.

 Hence, the function $\mathbf{S}_L$ generates a large portfolio with returns relative to the market portfolio given by:

\begin{eqnarray}
d \log( Z_{\xi}(t) / Z_{\mu}(t)) = d \log \mathrm{S}_L(\mu(t)) - \frac{1}{2} \xi_{p_t(m)}(t) d \Lambda_{\log \mu_{p_t(m)} - \log \mu_{p_t(m+1)}}(t), \label{eqn:largeport}
\end{eqnarray}
for all $t \in [0,T]$, a.s. Here all but the last local time term
drops out.

The generating function for the corresponding small portfolio is
given by:
\begin{eqnarray}
\mathbf{S}_S(x) = x_{p_t(m+1)} + \ldots + x_{p_t(n)},
\end{eqnarray} with
\begin{eqnarray}
d \log( Z_{\eta}(t) / Z_{\mu}(t)) = d \log \mathrm{S}_S(\mu(t)) +
\frac{1}{2} \eta_{p_t(m+1)}(t) d \Lambda_{\log \mu_{p_t(m)} - \log
\mu_{p_t(m+1)}}(t), \label{eqn:smallport}
\end{eqnarray}
 for $t \in [0,T]$. Here $\mathbf{S}_S(x)$ represents the relative
 capitalisation of a small stock index comprised of the $n-m$
 smallest stocks in the market.

Combining Equations (\ref{eqn:largeport}) and
(\ref{eqn:smallport}) we obtain the relative return of the small to large portfolios:
\begin{eqnarray}
d \log \left( { Z_{\eta}(t) \over Z_{\xi}(t) }\right)  &=& d \log
\left( {\mathrm{S}_S(\mu(t)) \over \mathrm{S}_L(\mu(t))} \right)
\\ && +\frac{1}{2} \left( {\xi_{p_t(m)}(t) + \eta_{p_t(m+1)}(t)}
\right) d \Lambda_{\log \mu_{p_t(m)} - \log \mu_{p_t(m+1)}}
\end{eqnarray}

Fernholz observed that if the relative capitalisation of the
small to large portfolio does not change significantly, then the
first term, dependent on $\mathbf{S}_S \over \mathrm{S}_L$, does
not change by much either. This implies that the relative returns
are dependent on the local time terms.

\section{Model-free discrete factorization of single \\ stock returns in a ranked market }

Throughout this section we consider a discrete time market, with
the $i^{th}$ stock modelled by it's market capitalisation $X_i(t)$
at time $t$ and the total market cap  denoted by $M(t) =
\sum\limits_{i=1}^N X_i(t)$. No assumptions are
required for the evolution of prices. It suffices that stocks can
be ranked by capitalisation to obtain the factorisation given in
Equation (~\ref{fact1}) for the case of no-dividends and Equation
(~\ref{fact2}) for a market with dividend paying stocks ~\cite{F1998}.

\vspace{0.3cm} At each time step and the weight of the  $i^{th}$
ranked stock is denoted as before as $ \mu_i(t) = {X_i(t) \over
M(t)}. $ In particular, we assume that these market weights are
conveniently  indexed in descending order according to size at
time $t$. Each time iteration brings price changes and possible
rank changes. Notationally this is captured  by letting
$\mu_{p_t(i)}(t)$ denote the weight of the $i^{th}$ ranked stock
at time $t$. The map $p_t$ denotes  a permutation which operates
on weights of ranked stocks and the subscript of $\mu$ points to
the rank at the time of measurement:

\begin{eqnarray}  \{ \mu_{1}(t_0) , \ldots , \mu_{n}(t_0) \} \stackrel{p_t}{ \longrightarrow} \left\{ \mu_{p_{t}(1)}(t), \ldots , \mu_{p_{t}(n)}(t) \right\}. \label{perm} \end{eqnarray}

Assuming no dividends,  after the iteration from time $t_0$ to
time $t$, the weight of the stock in  rank $i$ at time $t_0$,
$\mu_{i}(t_0)$, gets mapped to  the  weight of the stock  in rank
$i$ at time $t_1$, $\mu_{p_{t_1}(i)}(t_1)$. This is referred to as
the {\em  distribution component } of the transformation.
Meanwhile, the stock which was  at rank  $i$ at time $t_0$ may
occupy a new ranking at time $t_1$, which we can denote
$\mu_{p_{t_0}(i)}(t_1)$. The mapping from old rank to new of a
specific stock is referred to as the {\em time component}. This
can be graphically represented (as in \cite{F1998}).
\begin{displaymath}
    \xymatrix{
        \mu_{p_{t_0}(i)}(t_0) \ar[rd]^t \ar[d]_d & & \\
        \mu_{p_{t_1}(i)}(t_1) \ar[r]_r & \mu_{p_{t_0}(i)}(t_1)  & },
\end{displaymath}

The arrow labeled $d$ identifies the capital distribution part of the
transformation, while the $t$ labels the time component. The diagram is
completed with a transformation at time $t_1$ between ranks at that time,
which we refer to as the {\em rank component}.

\vspace{0.3cm} Now, still assuming no dividends, the log-return of
the $i^{th}$ stock is: \beq
\log\left(\frac{X_i(t_1)}{X_i(t_0)}\right), \eeq with a return on
the overall market portfolio for the same period computed as \beq
\log\left(\frac{M(t_1)}{M(t_0)}\right). \eeq Thus, computing
returns relative to the market one obtains: \beq
\log\left(\frac{X_i(t_1)}{X_i(t_0)}\right) -
\log\left(\frac{M(t_1)}{M(t_0)}\right)  =
\log\left(\frac{\mu_{p_{t_0}(i)}(t_1)}{\mu_{p_{t_0}(i)}(t_0)}\right).\eeq

From the triangle factorisation above, one obtains:
\begin{eqnarray}   \frac{\mu_{p_{t_0}(i)}(t_1)}{\mu_{p_{t_0}(i)}(t_0)} =\left( \frac{\mu_{p_{t_1}(i)}(t_1)}{\mu_{p_{t_0}(i)}(t_0)}\right)\left(\frac{\mu_{p_{t_0}(i)}(t_1)} {\mu_{p_{t_1}(i)}(t_1)}\right). \label{fact1} \end{eqnarray}

Taking logarithms in Equation (~\ref{fact1}), the right hand side is the sum of returns
attributed  to change in capital distribution plus a component due to change in rank:

\bea  \log \left({ \mu_{p_{t_0}(i)}(t_1) \over \mu_{p_{t_0}(i)}(t_0) }\right) = \log \left( { \mu_{p_{t_1}(i)}(t_1) \over \mu_{p_{t_0}(i)}(t_0) }\right) + \log \left({ \mu_{p_{t_0}(i)}(t_1) \over \mu_{p_{t_1}(i)}(t_1) }\right)  \eea


\vspace{0.3cm}

Dividends may be incorporated as the return ${\delta_{p_{t_0}(i)}(t_1)}$, measured
in portfolio weight of the stock which was in rank $i$ at time $t_0$, with an additional
branch to the factorisation:

\begin{displaymath}
    \xymatrix{
        \mu_{p_{t_0}(i)}(t_0) \ar[rd]^t \ar[d]_d & & \\
        \mu_{p_{t_1}(i)}(t_1) \ar[r]_r & \mu_{p_{t_0}(i)}(t_1) \ar[r]_{\delta} & \mu_{p_{t_0}(i)}(t_1) e^{\delta_{p_{t_0}(i)}(t_1)}}
\end{displaymath}

Thus, we have the following summary of maps:

\vspace{0.3cm}
\begin{tabular}{|c|c|}
  \hline
(Capital) Distribution component & $ \mu_{p_{t_0}(i)}(t_0) \longrightarrow_{d} \mu_{p_{t_1}(i)}(t_1) $ \\
  \hline
Rank components & $ \mu_{p_{t_1}(i)}(t_1) \longrightarrow_{r} \mu_{p_{t_0}(i)}(t_1) $ \\
  \hline
Time component & $ \mu_{p_{t_0}(i)}(t_0) \longrightarrow_{t} \mu_{p_{t_0}(i)}(t_1) $ \\
 \hline
Dividend component & $ \mu_{p_{t_0}(i)}(t_1) \rightarrow_{\delta} \mu_{p_{t_0}(i)}(t_1) e^{\delta_{p_{t}(i)}(t_1)}$ \\
\hline
\end{tabular}

\vspace{0.5cm}

The dividend rate for the market  from ${t_0}$ to ${t_1}$ is given by:

$$ \delta(t_1) = \ln \left({ \sum \mu_{p_{t_0}(i)}(t_1) e^{\delta_{p_{t_0}(i)}(t_1)}}\right), $$

where the dividend rate per stock for the time change between time ${t_0}$ and
${t_1}$ is denoted by $ \delta_{p_{t_0}(i)}(t_1).  $ Thus, the increase in value due
to the dividend rate, at time ${t_1}$, is

$$ \mu_{p_{t_0}(i)}(t_1) e^{\delta_{p_{t_0}(i)}(t_1)} $$

The dividend correction can, therefore, be computed as $
(\delta(t_1)+\delta_{p_{t_0}(i)}(t_1)) $ \ and the return
factorisation with dividends extends Equation (~\ref{fact1}) and
is given by:

\begin{eqnarray}  \log \left({ \mu_{p_{t_0}(i)}(t_1) \over
\mu_{p_{t_0}(i)}(t_0) }\right) = \log \left( {
\mu_{p_{t_1}(i)}(t_1) \over \mu_{p_{t_0}(i)}(t_0) }\right) + \log
\left({ \mu_{p_{t_0}(i)}(t_1) \over \mu_{p_{t_1}(i)}(t_1) }\right)
+ (\delta(t_1)+\delta_{p_{t_0}(i)}(t_1)). \label{fact2}
\end{eqnarray}


\subsection{Special-case proof of the size effect}\label{sizeprf}

The discrete time formulation is more simplistic but does not
clearly discriminate between the contributions due to drift and
changes in the relative capitalisation of stocks. However, it does
afford some insight into how the mechanical construction is
independent of the price process. Clearly if one has two mutually
exclusive portfolios built on ranking stocks, there will
cross-overs from the one portfolio to the other due to price
volatility. In particular, outperformance can occur as statistical
effect without in any way being risk premium

Consider a large cap portfolio, $Z_L$ of the biggest $m$ stocks
and a small portfolio, $Z_S$,  comprising the rest of the market:

\begin{eqnarray}
Z_L(t_0) &=& X_{p_{t_0}(1)}(t_0)  + \ldots + X_{p_{t_0}(m)}(t_0), \label{eqn:valuelarge} \\
Z_S(t_0) &=& X_{p_{t_0}(m+1)}(t_0)  + \ldots +
X_{p_{t_0}(n)}(t_0). \label{eqn:valuesmall}
\end{eqnarray}

At some later time $t_1$ the return for the $i$-th stock is
$r_i(t_0)  = \log(X_i(t_1) /X_i(t_0) )$ so that the return of the
large portfolio and small portfolio are
\begin{eqnarray}
r_L(t_1) &=& \log (X_{p_{t_0}(1)}(t_1) + \ldots + X_{p_{t_0}(m)}(t_1) / Z_L(t_0) ) \label{eqn:returnlarge}\\
r_S(t_1) &=& \log (X_{p_{t_0}(m+1)}(t_1) + \ldots +
X_{p_{t_0}(n)}(t_1) / Z_S(t_0) ), \label{eqn:returnsmall}
\end{eqnarray}

where the portfolio returns are computed before rebalancing
according to new ranks. The portfolios are then reconstituted at
time $t_1$. We now review the result  that if the relative ratios
remain constant then the return of the small portfolio is greater
than the return of the large portfolio.

\vspace{0.3cm}First observe that if the capitalisation ratios of
the rebalanced portfolios do not change \cite{F1998} then
\begin{eqnarray}
{Z_L(t_1) \over Z_L(t_0)} = {Z_S(t_1) \over Z_S(t_0)}.
\label{eqn:values}
\end{eqnarray}

Now the large portfolio is comprised of the $m$ largest stocks at
each time and hence:

\beq Z_L(t_1) \geq X_{p_{t_0}(1)}(t_1) + \ldots +
X_{p_{t_0}(m)}(t_1) \eeq with equality if and only if
$\{p_{t_1}(1),\dots,p_{t_1}(m)\} =
\{p_{t_0}(1),\dots,p_{t_0}(m)\}$.

\vspace{0.3cm} Similarly $Z_S(t_1) \leq X_{p_{t_0}(m+1)}(t_1) +
\ldots + X_{p_{t_0}(n)}(t_1)$.

It follows from Equations (\ref{eqn:returnlarge})
(\ref{eqn:returnsmall}) respectively, that
\begin{eqnarray}
r_L(t_1) &\leq& \log( Z_L(t_1) / Z_L(t_0)) \\
r_S(t_1) &\geq& \log( Z_S(t_1) / Z_S(t_0)).
\end{eqnarray}

Thus, by Equation (\ref{eqn:values}),  $r^{t_1}_L \leq r^{t_1}_S$.

\vspace{0.3cm}Heuristically, the effect of rebalancing (in
response to cross-overs in ranking) is that at each time companies
performing worse (or that have had capitalisation reduced) get
dropped into the small stock portfolio and stocks that have done
well (or that have seen an increase in market capitalisation) in
the small cap portfolio get bumped up. However, the computation of
returns $r_L(t_1)$ and $r_S(t_1)$ is based on ranks before
rebalancing, so that $r_L(t_1)$ catches the poor performance while
$r_S(t_1)$ locks in returns from the better performers. All other
things being equal, higher volatility in a continuous model
implies that portfolios would need to be rebalanced more often.

\section{Factorising returns of portfolios}

Summarising results from \cite{F1998}, we recall the discrete time
factorisation of returns at the portfolio level. 
Consider a portfolio $V$ with stock positions given by the coefficients $v_i$:

\beq  V(t)  = \sum_{i=1}^n v_{i} X_{i}(t). \eeq

\newpage
The portfolio weights are given by

\beq  \pi_{i}(t)  = \frac{v_{i} X_{i}(t)}{ V(t)} \eeq

and the weight ratios are computed by

\beq \frac{\pi_{i}(t)}{ \mu_{i}(t)}  = \frac{v_{i} M(t)}{ V(t)}. \eeq

The portfolio generating function $(S(x,v))$ for $V$ satisfies

\begin{eqnarray}
\mathbf{S}(x(t),v) = \sum_{i=1}^n v_{i}x_{i}(t).
\end{eqnarray}

Portfolio factorisation via ranking incorporates the following expression
which is referred to as the  {\em rank sensitivity function} $(R(x,v)) $ :

\begin{eqnarray} \mathbf{R}(x(t),v) = {{\sum_{i=1}^n v_{i} x_{i}(t)} \over {\sum_{i=1}^n v_i x_{p_{t}(i)}(t)}}
\end{eqnarray}

This term captures the effect of changes in rank and $\log R$ corresponds to
the drift function in the continuous time model.

\vspace{0.3cm}
Proceeding as in Section 3 with the factor diagrams, returns relative to the market are computed as:

\beq \log \left( {V(t_1) \over V(t_0)} \right) - \log \left( {M(t_1) \over
M(t_0)} \right) = \log \left(  {\bf S}(\mu(t_1)) \over {\bf S}(\mu(t_0)) \right)
+ \log \left( {\bf R}(\mu(t_1)) \right) +
(\delta_v(t_1) - \delta(t_1)).   \eeq

Similarly, the performance of one portfolio relative to another portfolio is given by:

\beq  \log \left( {V(t_1) \over V(t_0)} \right) - \log \left( {W(t_1) \over
W(t_0)} \right) = \log \left( {{\bf S}_v(\mu(t_1)) \over {\bf S}_w(\mu(t_1))} \right)
+ \log \left( {{\bf R}_v(\mu(t_1))\over {\bf R}_w(\mu(t_1))} \right) +
(\delta_v(t_1) - \delta_w(t_1)) .  \eeq

As before the first contribution on the right is due to the distributional component,
the second the rank component (analogous to the drift component) and the last component
is that of the relative return of dividends.

\newpage
\section{Discrete approximations for the continuous time SPT model}

We recall Equation (\ref{eqn:312F2002}):

\begin{eqnarray*}
d \log( Z_{\pi}(t) / Z_{\mu}(t)) = d \log \mathrm{S}(\mu(t)) + d
\Theta(t).
\end{eqnarray*}

The first term on the right hand side of  Equation
(\ref{eqn:312F2002}) is referred to as the distributional
component. For portfolios which incorporate rank processes, this
can be approximated as
\begin{eqnarray}
d \log (S(\mu(t_0))) \approx \log \left({{S(\mu_{p_{t_1}(1)}(t_1),
\ldots, \mu_{p_{t_1}(m)}(t_1))} \over S(\mu_{p_{t_0}(1)}(t_0),
\ldots, \mu_{p_{t_0}(m)}(t_0))} \right).
\end{eqnarray}

The drift term in Equation (\ref{eqn:312F2002}) includes a smooth
contribution and the local-time contribution. The local-time
contribution can be approximated by:
\begin{equation}
d L_{\pi}(t_0) \approx \log \left(
{S(\mu_{p_{t_0}(1)}(t_1),\ldots,\mu_{p_{t_0}(n)}(t_1)) \over
S(\mu_{p_{t_1}(1)}(t_1),\ldots,\mu_{p_{t_1}(n)}(t_1))}\right),
\end{equation}
where the time increment from time $t$ to time $t+1$ is
sufficiently small. This corresponds to the rank component in the
discrete time factorisation.

\section{Main results: components of market \\ behaviour}

Monthly data for stocks listed on the JSE mainboard between Oct
1994 and Jun 2007 from Thomson Reuters Datastream is used to construct portfolios based on rank
and  the latter are then analysed with respect to the evolution of
distribution, rank, total (time) and dividend components. We first
split the stock universe into portfolios comprised of shares which
are ranked by market weight as the top 50 (Large), the top 51-100
(Mid) and top 101-250 (Small). We also examined an alternative
split into  the top 10 for the Large, the next 30 stocks for the Mid
and stocks ranked  41-165  for the Small cap portfolio.  Aggregate results for the two decompositions are tabulated in Section ~\ref{aggr}. In the rest of this section we discuss only the first market partitioning.  We also 
consider decomposition of returns of so-called value stocks. We
construct a Value portfolio comprised of the top 50 stocks ranked
according to {\em book-value-to-price}  and perform the corresponding
analysis.

\begin{figure}
\centering
\includegraphics[width=12cm]{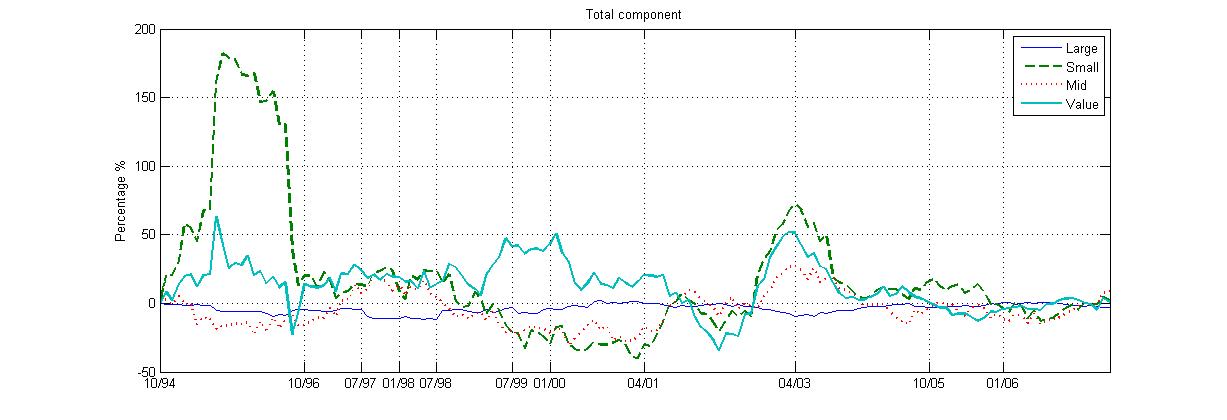}
\caption{Total component}\label{total}
\includegraphics[width=12cm]{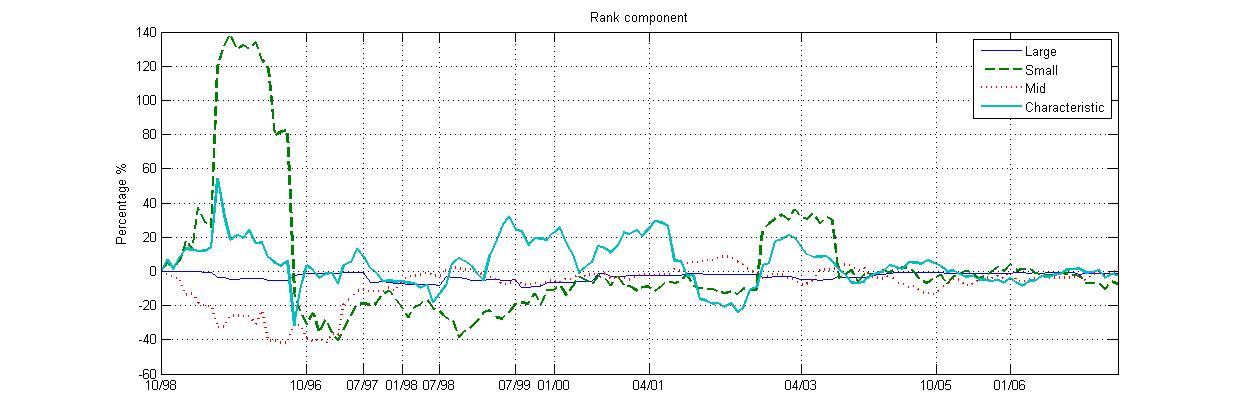}
\caption{Rank component}\label{rank}
\includegraphics[width=12cm]{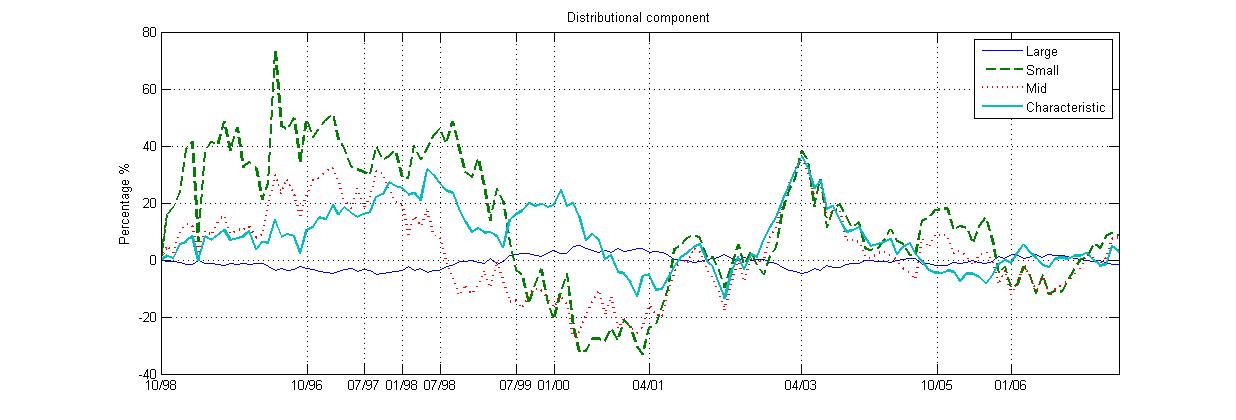}
\caption{Distribution component as rolling annual returns.}\label{dist}
\includegraphics[width=12cm]{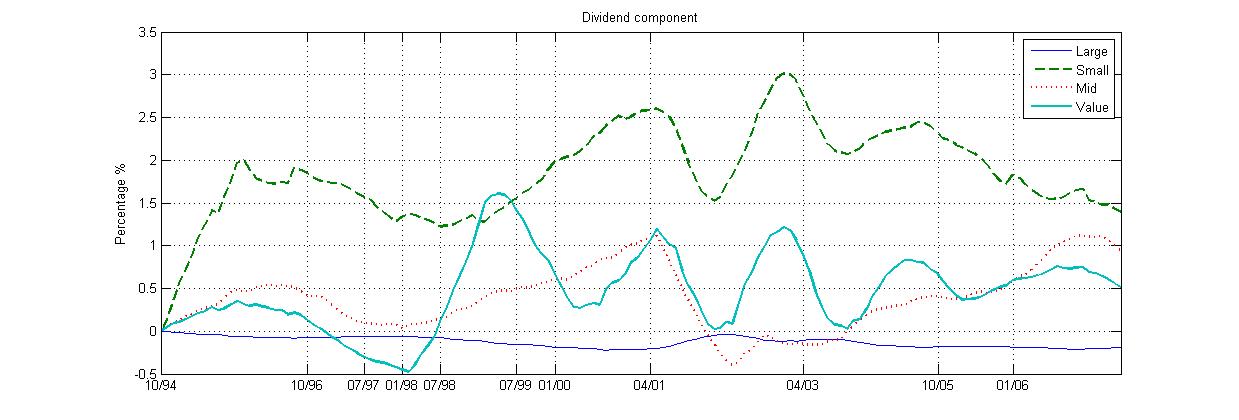}
\caption{Dividend component}\label{divi}
\end{figure}

\vspace{0.3cm}
Figures ~\ref{total},  ~\ref{rank}, ~\ref{dist} and  ~\ref{divi} present the results
for the four components considered for small, mid, large and value portfolios. We observe
that components are far from stable over time. Numerous exogenous impacts during the
period under consideration also cause the market to transition through various regimes over
time. For our discussion we identify the following key market sentiment triggers:

\begin{itemize}
\item The first universal franchise elections in SA in April 1994.
\item The ZAR  currency crises of Feb and Oct 1996.
\item The Asian crisis of 1997-1998 commencing the Thai Baht crisis in May 1997.
\item The ZAR-USD crisis April-Aug 1998.
\item The Russian GKO bond default Aug 1998.
\item The Dot-com bubble which peaked in March 2000.
\item The ZAR-USD crash of Dec 2001.
\end{itemize}

The early part of the investigation period saw the phasing out of financial rand and while
1996-2001 saw 3 significant currency crises. While the 1998 currency crisis can at least be
 understood as a contagion effect of the Asian crisis, the 1996 and 2001 crisis have been
 explained is some parts of the literature as speculative attacks and/or herding phenomenon
 and are not necessarily the response to negative conditions with the market.
 The exchange rate has therefore been a natural choice to be included as a macroeconomic
 risk factor in Ross's arbitrage pricing theory.

\begin{figure}
\includegraphics[width=12cm]{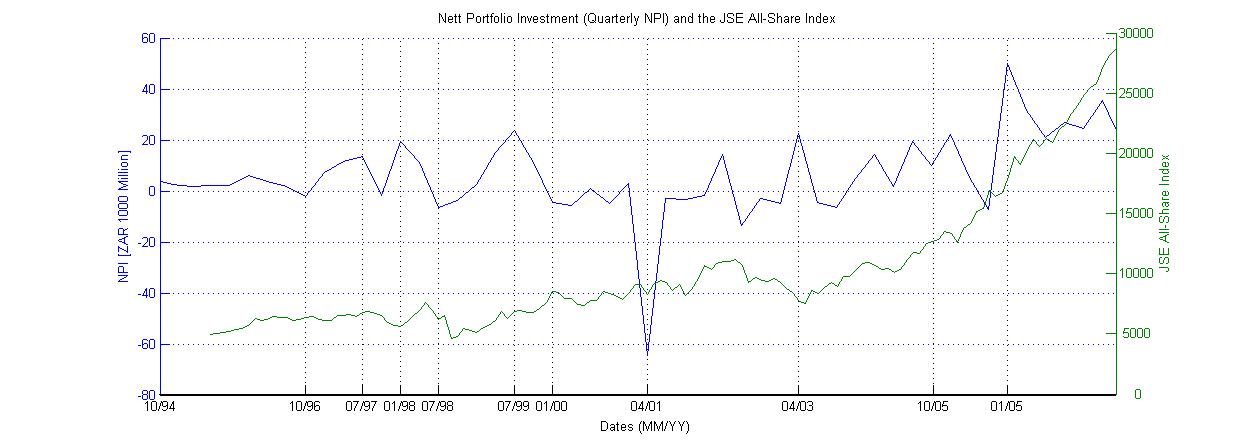}
\caption{Net portfolio investment (NPI) flows to South Africa, 1994-2007 [Source: South African Reserve Bank] and JSE All-share Index}\label{fpi}
\end{figure}

\vspace{0.3cm} From the graphic information  we identify
lag and lead connections between net portfolio investment
(NPI)  and the performance of returns in the market. This contrasts with the findings of \cite{JF2011}, where  only a lag impact of weekly return on NPI was identifiable between Jan 2002 and Dec 2006. We note that the difference in findings could  be partially attributed to differing periods considered. Figure ~\ref{fpi} depicts
quarterly cross-border portfolio flows for the period under
investigation.  We identify nine significant quarters within 1994-2007 for
which there were obvious correlation effects between NPI levels and
market behaviour with respect to overall performance of rank-based
sectors as well as capital distribution components, rank
volatility and dividend policies. We note that the dividend component of the Small portfolio outperforms all the other portfolios for almost the entire period of 1994-2007.  The variation in the Total component for Small portfolios can therefore not be explained by dividend policies along. The distribution components of Small,  Mid and Value portfolios are significant from 1994 through to 1999, implying that stocks are moving up in rank through that time. 

\begin{enumerate}
\item{\bf1994:} The start  of the period coincided with a huge increase in NPI.
The Small portfolio was most dramatically affected, followed by the
 Value portfolio. Here we see a dramatic increase in investment in these parts of the market,
 followed by equally dramatic declines in returns by 1996, correlated to the ZAR
 currency crisis of that period and highlighting the fragility of small caps under volatile NPI. The rank component of the Small portfolio dominates returns initially and then collapses. A large rank component indicates that as stocks move up in 
rank there is a large difference between their new market weights and the weights of stocks occupying their former ranks.  

\item{\bf Negative NPI in 4th quarter of 1996:} This was the first NPI outflow after 1994.
Here negative NPI lags the dramatic 1996 ZAR currency depreciation. The latter cannot be attributed to any obvious macroeconomic signals such as increase in inflation (which was in fact decreasing at that time) or increase in unemployment, but instead seems to have been triggered by negative sentiment on the  new governments performance \cite{DMVW}.  The ZAR crisis can therefore explain both the reduction in NPI and the dramatic fall-off in the Small and Value
portfolios. This period is also the beginning of a steady improvement in Mid cap stocks, some of which were Small caps in the preceding period. 

\item{\bf Peak in NPI by 3rd quarter of 1997:} This lags a peak in the Value portfolio that year
 but leads  peaks in the performance of Small and Mid portfolios and leads peaks in the
 distributional and rank components of Small, Mid and Value portfolios.

\item{\bf Peak in NPI in 1st quarter of 1998:} This leads an increase in the performance of the Mid portfolio.

\item{\bf Negative NPI in 3rd quarter of 1998:} This leads a fall in the Small portfolio

\item {\bf Peak in NPI in 3rd quarter of 1999:} This lags a peak in Value  and all its
components, but leads an improvement in the Small portfolio.

\item{\bf Negative NPI in 1st quarter of 2000:} This leads a fall in the Value portfolio.

\item{\bf Plummet in NPI in 2nd quarter of 2001:} This can be attributed to the collapse
of the Dot-com bubble. It lagged negative trends in Small and Mid portfolios and
components thereof, but preceded negative performance of the Value portfolio.

\item{\bf Significant peak by end of 1st quarter of 2003:} This coincided with peaks in all portfolios and their distributional components.

\item{\bf Significant NPI drop by 4th quarter of 2005:} This can be attributed to a big drop in US equity values at that time. 

\item{\bf Dramatic increase in NPI at the start of 2006:} This has limited discernable impact as investment in the JSE from with in SA increased dramatically. The increase in NPI and overall market value led to an improved performance of the Small portfolio. 
\end{enumerate}

%
%



\subsection{Relative performances of rank-based portfolios and
components}

\begin{figure}
\centering
\includegraphics[width=12cm]{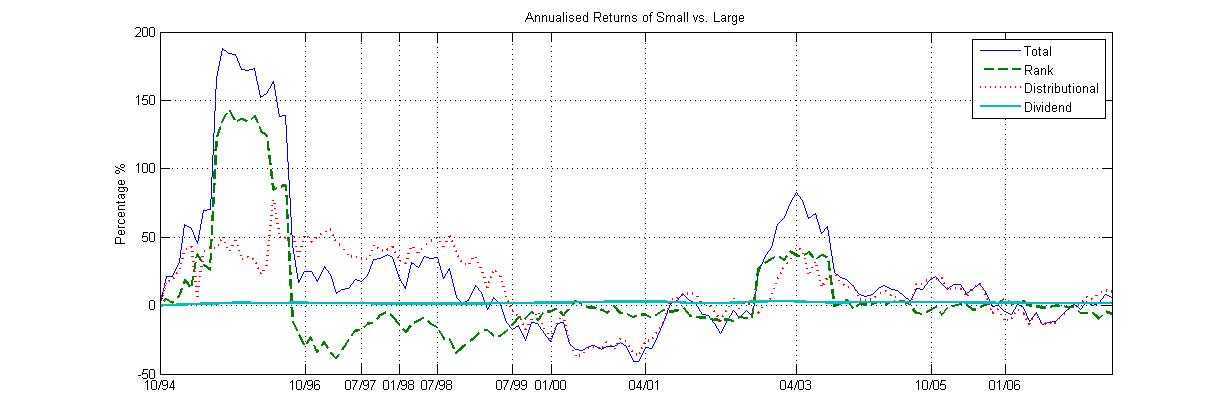}
\caption{Components of Small relative to Large return ratios}\label{sl}
\includegraphics[width=12cm]{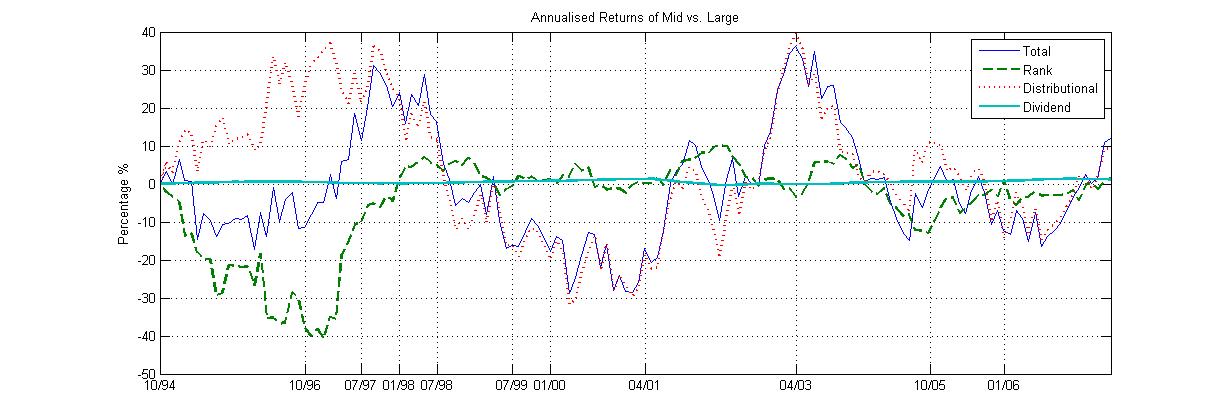}
\caption{Components of Mid relative to Large return ratios}\label{ml}
\includegraphics[width=12cm]{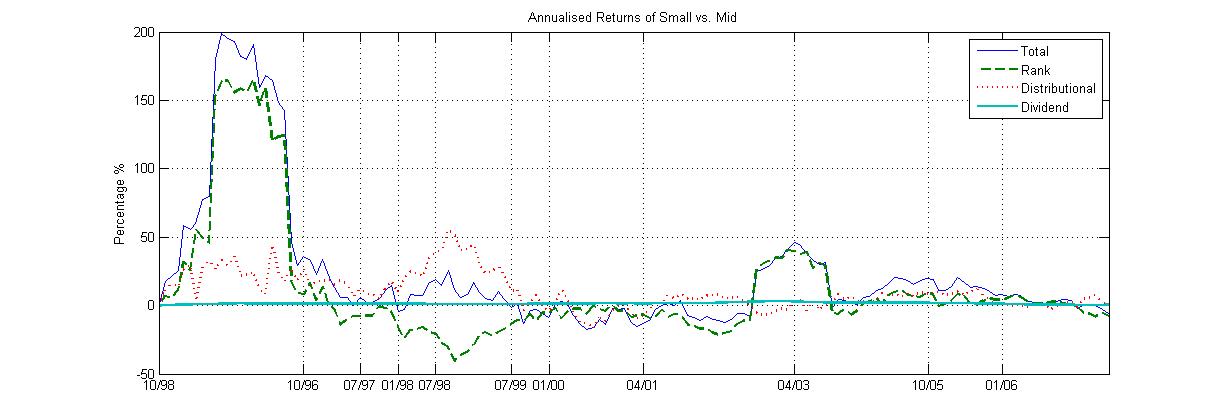}
\caption{Components of Small relative to Mid return ratios }\label{sm}
\includegraphics[width=12cm]{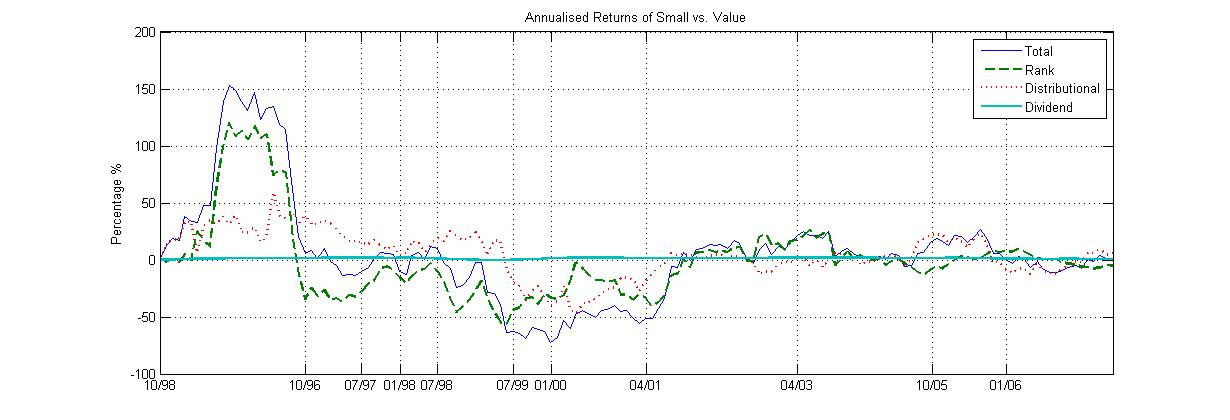}
\caption{Components of Small relative to Value return ratios}\label{sv}
\end{figure}

The next sequence of figures illustrate the relative portfolio performances, namely Small to Large,
Mid to Large, Small to Mid and Small to Value. As before there are no immediately
obvious trends. Instead notable regime changes can be understood in the context of
exogenous impacts, including significant changes in NPI flows. As before, the beginning of the date range sees the Small cap portfolio and its components dominating the market. Next, Figure ~\ref{ml} corroborates that Mid cap stocks start to outperform other parts of the market after the 1996 currency crisis and brief NPI reversall. Large porfolios begin to fare better relative to Small and Mid after the Asian crisis and up to the Dotcom bubble. After the market resumes aggregate growth in 2003, triggered by an initial jump in NPI in the second quarter of that year, growth takes place in all parts of the market. The small contributions of the SPT components in that time suggests parallel growth across the market, as opposed to investors changing preference from one section of the market to another.  

%
%
%

\newpage 
\subsection{Approximating the Local times}

Local time components in the continuous time model are similar to  rank
components under the discrete equity return factorisation. However local times are strictly positive and large local times in a section of the market imply that stocks are changing rank rapidly, moving either up or down. From Equations
(\ref{eqn:largewts}) and (\ref{eqn:largeport}) we
have:

\begin{eqnarray}
d \log \left( {Z_{\xi}(t) \over Z_{\mu}(t)} \right)
&=& d \log ( \mu_{p_{t}(1)}(t) + \ldots + \mu_{p_{t}(k)}(t))
 \\ & & - \frac{1}{2} {\mu_{p_{t}(k)}(t) \over (\mu_{p_{t}(1)}(t) + \ldots + \mu_{p_{t}(k)}(t))} d
 \Lambda_{\log \mu_{p_{t}(k)} - \log \mu_{p_{t}(k+1)}}(t)
 \end{eqnarray}

One can compute the local times for different values of $k$  from the evolution
of the portfolio, $Z(t)$, and the
reconstituted portfolio using the generating function $S$,  by integrating
the differential:

\begin{eqnarray}
 d \Lambda_{\log \mu_{p_{t_0}(k)} - \log \mu_{p_{t_0}(k+1)}}(t) \approx 2 \left( {\mu^{t_0}_{p_{t_0}(1)}+\ldots+\mu^{t_0}_{p_{t_0}(k)} \over \mu_{p_{t_0}(k)}^{t_0}} \right) \\
 \times \left[ {\log \left({{\mu_{p_{t_1}(1)}^{t_1}+\ldots+\mu_{p_{t_1}(k)}^{t_1}} \over \mu_{p_{t_0}(1)}^{t_0} + \ldots + \mu_{p_{t_0}(k)}^{t_0}}\right)} - \log \left({X_{p_{t_0}(1)}^{t_1} + \ldots + X_{p_{t_0}(k)}^{t_1} \over X_{p_{t_0}(1)}^{t_0} + \ldots + X_{p_{t_0}(k)}^{t_0}} \right) - \log \left( {M^{t_1} \over M^{t_0}} \right) \right].
\end{eqnarray}
Here $\mu_i$ are the weights in the full market portfolio $\mu$
with value $Z_{\mu}$ and $\xi_i$ are the weights in the large
portfolio $\xi$ for the first $k $ ranked weights and with value
$Z_{\xi}$.

\begin{figure}
\centering
\includegraphics[width=10cm]{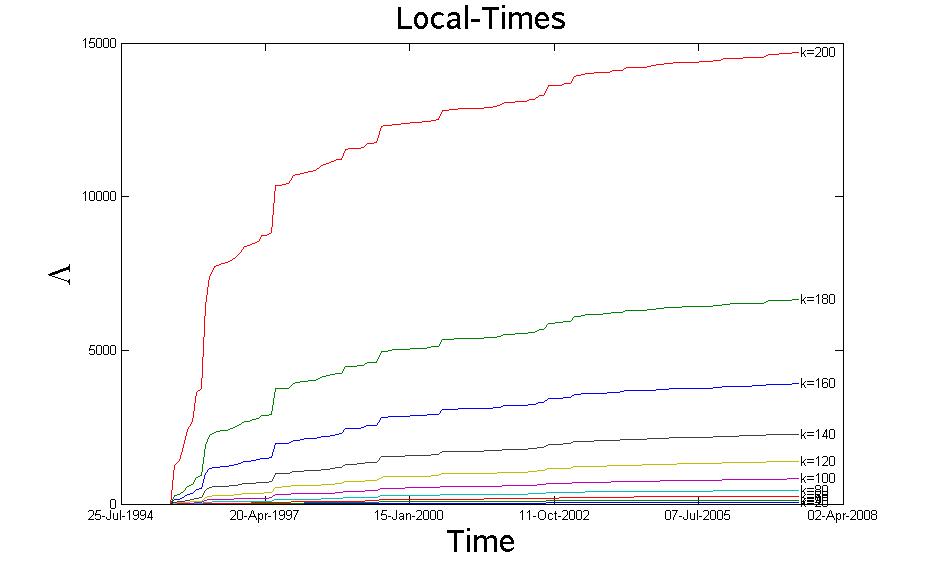}
\caption{Local times for the largest 200 stocks on the JSE
(1994-2007) }\label{loct2}
\includegraphics[width=10cm]{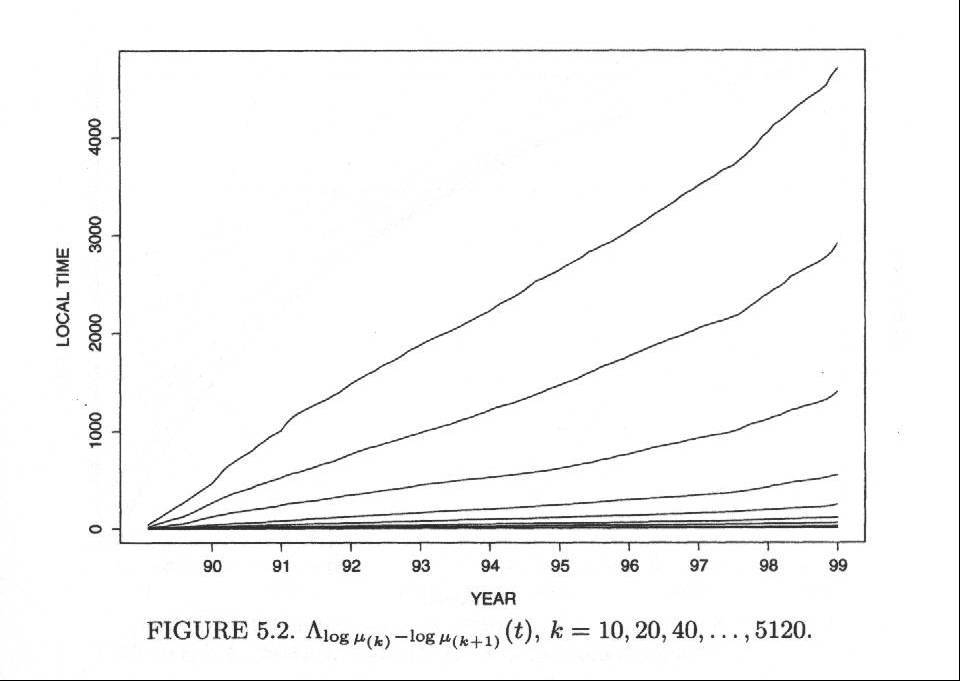}
\caption{Local times for monthly data for the largest 5120 stocks
listed in the CRSP database of US equities, 1 Jan 1990 - 31 Dec
1999\cite{F2002}. }\label{loct1}
\end{figure}

In Figure ~\ref{loct2} it is clear that the local times of small cap stocks see dramatic
increments at the start of the period under investigation. This is consistent with the
dramatic increase in the rank component of Small cap portfolios at that time. It suggests
that relatively high NPI inflows (for that time)  contributed to lots of volatility in market
 ranks at the Small cap end of the market.   By the end of the period increments in local
 times flatten off throughout the market, signalling more stable market ranks. A very different
  situation is described in Figure ~\ref{loct1} for  the local times for  US equities  between
  1990 and 1999 \cite{F2002}, where local times increase (almost)  linearly. The latter implies fairly
  constant rates in the change of stock ranks   through time.

\newpage

\subsection{Aggregate comparisons}\label{aggr}
Tables 1-3 document
overall relative performances and contributions of components. The monthly data is
decimated to consider quarterly, semi-annual and annual performances as well.
We report findings for two distinct partitions of the overall stock universe,
namely  the case when the Large portfolio consists of the Top 10 stocks and the
Small portfolio is comprised of the remaining shares in the Top 250 and the
case when Large consists of the Top 40 and Small is made up of the complement of stocks in the Top 165.

\normalsize
\begin{table}
\begin{footnotesize}
\begin{tabular}{cccccc}
\hline
Portfolio & Total & Rank & Distributional & Dividend \\
\hline
\multicolumn{5}{c}{m=40,n=165}\\
\hline
Small/Large &  2.1 (15.8) & -2.3 (8.3) &  3.7 (15.6) &  0.7 (0.1) \\
Small/Value & -12.4 (21.1) & -9.0 (20.5) & -3.4 (10.6) &  0.0 (0.2) \\
\hline
\multicolumn{5}{c}{m=10,n=250}\\
\hline
Small/Large &  3.6 (16.7) &  3.3 (7.0) &  0.1 (14.9) &  0.2 (0.2) \\
Small/Value & -1.2 (86.1) & 21.1 (94.0) & -22.9 (38.5) &  0.6 (0.6) \\
\hline
\end{tabular}
\end{footnotesize}
  \caption{We tabulate the components of Small and Large portfolios for 2 different
  partitions of the stock universe.  We observe that the distributional component is
  dominant when the Large portfolio is comprised of the top 40 stocks  and the Small
  portfolio is made up of the remaining stocks in the the top 165 by market capitalisation.
This is a counter example to the argument that the rank contribution can be used to explain
  the difference of performance between Large and Small cap portfolios (which can be
  proved for the special case when ratios of capitalisations of Large to Small are unchanged through time).
  We find that if the Large portfolio is comprised of the top 10 stocks, with the
  remaining stocks in the top 250 making up the Small portfolio, then the rank contribution is dominant
  and the distributional component is much less significant.}\label{tab:T1}
\end{table}
\normalsize

\normalsize
\begin{table}
\begin{footnotesize}
\begin{tabular}{cccccc}
\hline
Component & Portfolio & Monthly & Quarterly & Semi-annual & Annual \\
\hline
Total & Large & -0.328 (0.9) & -0.890 (1.6) & -1.759 (2.1) & -3.311 (3.1) \\
Total & Small & 1.546 (9.9) & 2.570 (9.3) & 5.480 (13.7) & 14.073 (23.6) \\
Total & Mid & -0.349 (3.9) & 0.339 (6.5) & 1.160 (8.5) & 2.349 (11.7) \\
Total & Value & 1.060 (5.9) & 3.805 (6.8) & 6.688 (13.1) & 14.550 (14.6) \\
\hline
Distributional & Large & -0.011 (0.0) & -0.031 (0.0) & -0.060 (0.0) & -0.118 (0.1) \\
Distributional & Small & 0.157 (0.1) & 0.475 (0.2) & 0.969 (0.5) & 2.137 (2.4) \\
Distributional & Mid & 0.035 (0.0) & 0.098 (0.1) & 0.184 (0.3) & 0.263 (0.5) \\
Distributional & Value & 0.041 (0.1) & 0.123 (0.2) & 0.232 (0.4) & 0.386 (0.8) \\
\hline
Rank & Large & -0.060 (0.6) & -0.167 (1.1) & -0.345 (1.4) & -0.641 (2.2) \\
Rank & Small & 1.104 (6.3) & 2.786 (8.5) & 5.652 (12.4) & 10.497 (20.0) \\
Rank & Mid & 0.271 (4.0) & 0.706 (7.2) & 1.379 (9.2) & 2.560 (14.7) \\
Rank & Value & 0.670 (2.8) & 1.485 (4.5) & 2.014 (5.3) & 3.917 (8.8) \\
\hline
Dividend & Large & -0.258 (0.7) & -0.692 (1.2) & -1.354 (1.6) & -2.551 (2.2) \\
Dividend & Small & 0.285 (9.1) & -0.691 (6.5) & -1.142 (7.8) & 1.439 (10.8) \\
Dividend & Mid & -0.655 (2.4) & -0.465 (3.9) & -0.402 (5.3) & -0.475 (6.5) \\
Dividend & Value & 0.349 (5.1) & 2.197 (5.9) & 4.442 (9.7) & 10.248 (10.4) \\
\hline
\end{tabular}
\end{footnotesize}
  \caption{The total, distributional, rank and dividend components of the return
  factorisation relative to the market portfolio are tabulated. Figures are reported for
  the case when the Large portfolio consists of the Top 10 stocks (Top 40 in brackets) and
  the Small portfolio is comprised of the remaining shares in the Top 250 (Top 165). All
  returns are average annualised returns at
  the different sampling frequencies. The sampling frequencies where constructed from the
  original data by decimating to the required sampling rate. }\label{tab:T2}
\end{table}
\normalsize

\normalsize
\begin{table}
\begin{footnotesize}
\begin{tabular}{cccccc}
\hline
Portfolio & Component & Monthly & Quarterly & Semi-annual & Annual \\
\hline
Small/Large & Total & 22.5 (124.316843) & 13.8 (40.781762) & 14.5 (29.780399) & 17.4 (25.820635) \\
Small/Large & Rank &  6.5 (111.779406) &  0.0 (26.856601) &  0.4 (15.868101) &  4.0 (10.856648) \\
Small/Large & Distributional & 14.0 (81.085951) & 11.8 (37.948269) & 12.0 (27.340062) & 11.1 (22.079728) \\
Small/Large & Dividend &  2.0 (0.738672) &  2.0 (0.760598) &  2.1 (1.007544) &  2.3 (2.469674) \\
\hline
Small/Value & Total &  5.8 (96.082530) & -4.9 (38.792192) & -2.4 (23.799369) & -0.5 (17.555168) \\
Small/Value & Rank & -0.8 (91.543140) & -11.6 (30.244967) & -11.2 (22.664692) & -8.8 (13.881393) \\
Small/Value & Distributional &  5.2 (65.417202) &  5.2 (32.408844) &  7.3 (18.375326) &  6.6 (13.060872) \\
Small/Value & Dividend &  1.4 (0.764702) &  1.4 (0.751146) &  1.5 (0.843939) &  1.8 (2.083789) \\
\hline
\end{tabular}
\end{footnotesize}
  \caption{Relative portfolio performances are considered for Small vs. Large and Small
  vs. Value. Figures are reported for the case when the Large portfolio consists of the Top 10 stocks (Top 40 in brackets) and the Small portfolio is comprised of the remaining shares in the Top 250 (Top 165). Returns computed based on annualised values for each sampling frequency. Data at lower
  sampling frequencies was obtained by decimation of the original monthly data. The
  distributional component is more stable
  for Small vs. Large. The rank component is significant for Small vs. Large on shorter
  time-scales while for longer periods this holds for Small vs.Value portfolios (with a sign change).
  Small outperforms Large on all
  sampling rates, but only outperforms Value on the monthly time-scales. The rank
  contribution for Small portfolio contributes less on aggregate to total returns than for the Value
  portfolio. }\label{tab:T3}
\end{table}
\normalsize

\newpage

\section{Conclusion}

We applied Fernholz's stochastic portfolio factorisation to
size-ranked portfolios of stocks listed on the JSE between 1994
and 2007 to tease apart rank, distributional and dividend
components of total returns. We identify notable regime changes in the sentiment of investors through changes in dominating portfolios and portfolio components  which correspond to changes in NPI levels rather than overall market performance. 

The period under investigation
coincided with several exogenous market shocks, including currency
crises, which had varying impacts on different parts of the
market.
We highlight that outperformance of size-based portfolios can 
be attributed to capital allocations under significant NPI
inflows. This is consistent with the perspective that, in the
absence of any other information, investors pick Small cap or Mid cap portfolios 
in anticipation of capacity for growth.  Figures ~\ref{total} and
~\ref{rank} highlight that Small caps are also particularly
vulnerable to NPI flow reversals. Our investigation confirms that NPI in a given market is often exogenously determined,  with corresponding (lead) impacts on the market considered as opposed to being driven by market performance as a lagging factor. 

\section*{Acknowledgements}

This research was partially funded by a Carnegie Foundation
research grant administered by the University of the Witwatersrand
and an National Research Foundation NPYY Grant [Number 74223].


\begin{thebibliography}{00}


\bibitem{Ashman}
Ashman, S., Fine, B., Newman, S., (2011) Amnesty International? The Nature,
Scale and Impact of Capital Flight from South Africa, {\em Journal of Southern African Studies}, 37:01, 7-25


\bibitem{Ball}
Ball, R., (1978),  Anomalies in Relationships Between Securities'
Yields and Yield-Surrogates, {\em Journal of Financial Economics}
6, 103-26.



\bibitem{Banz}
Banz, R., (1981) The Relationship Between Return and Market Value of Common Stock,
{\em Journal of Financial Economics}, 9,  3-18.

\bibitem{Basu}
Basu, S., (1977),  Investment Performance of Common Stocks in
Relation to their Price-Earnings Ratio: A Test of the Efficient
Market Hypothesis, {\em Journal of Finance}, 32, June, 663-682.


\bibitem{Lux}
Colander, D., F\"{o}llmer, H.,  Haas, A., Goldberg, M., Juselius, K., 
Kirman, A., Lux, T., Sloth, B., (2009) The Financial Crisis and the Systemic Failure of Academic Economics, Kiel Working Paper 1489

\bibitem{DMVW}
Diamond, A., Manning, S., Vasquez, J., Whitaker, E., (2003)
South African Trade-Offs among Depreciation, Inflation, and Unemployment, Working paper

\bibitem{Eich}
Eichgreen, B., (2013)
Currency War on International Policy Coordination? Working paper

\bibitem{FF1992} Fama, E., French, K., (1992) The cross-section
of expected returns, \emph{Journal of Finance}, 47, 427-465

\bibitem{FF1996} Fama, E., French, K., (1996) Multifactor explanations of asset
pricing anomalies, \emph{Journal of Finance}, 51, 55-87

\bibitem{FS1982} Fernholz, R., Shay, B., (1982) Stochastic Portfolio Theory and Stock Market Equilibrium
\emph{{Journal of Finance}}, 37, 2, 615-624

\bibitem{F1998} Fernholz, R., (1998) Cross-overs, dividends, and the size
effect, \emph{Financial Analyst Journal} 54(3) 73-78

\bibitem{F2001} Fernholz, R., (2001) Equity portfolios generated by functions of
ranked market weights, \emph{Finance and Stochastics} 5, 469-486

\bibitem{F2002} Fernholz, R., \emph{Stochastic Portfolio Theory},
Springer - Applications of Mathematics, 2002


\bibitem{JF2011}
French, J.J., (2011) The dynamic interaction between foreign equity flows and returns: Evidence from the Johannesburg Stock Exchange, {\em International Journal of Business and Finance Research}, Vol. 5, No. 4, 45-56


\bibitem{GW2008}Gebbie, T., Wilcox, D., (2008) Faking value and Size, {Collective Insights}, Special Issue (Myth-busters) of Financial Mail, Nov 2009.

\bibitem{Lintner}
Lintner, J. (1965), The valuation of risk assets and the selection
of risky investments in stock portfolios and capital budgets, {\em
Review of Economics and Statistics}, 47 (1), 13-37.

\bibitem{McC2012} McCauley, R., (2012) {Risk-on/Risk-off, capital flows, leverage and safe assets}, \emph{BIS Working Papers}, N0. 382

\bibitem{Mohamed}
Mohamed, S., The impact of international capital flows on the South Africa
economy since the end of Apartheid,
http://www.policyinnovations.org/ideas/policy\_library/data/01386/

\bibitem{Mossin}
Mossin, J. (1966),  Equilibrium in a Capital Asset Market, {\em
Econometrica}, Vol. 34, No. 4, pp. 768-783.

\bibitem{RR2003} Rogoff, K., Reinhart, C., (2003) FDI to Africa: The Role  of Price Stability and Currency Instability, \emph{ IMF Working Paper}, WP/03/10

\bibitem{Ross}
Ross, S., (1976) The arbitrage theory of capital asset pricing,
{\em Journal of Economic Theory} 13, 341-60.

\bibitem{Sharp}
Sharpe, W. F.,  (1964) Capital asset prices: A theory of market
equilibrium under conditions of risk, {\em Journal of Finance}, 19
(3), 425-442

\bibitem{SvH}
Svetlova, E., van Elst, H., (2012) How is non-knowledge represented in economic theory, arXiv:1209.2204v1 [q-fin.GN]

\bibitem{Treynor}
Treynor, J.L. (1962),  Toward a Theory of Market Value of Risky
Assets,  Unpublished manuscript, final version  published in 1999,
in {\em Asset Pricing and Portfolio Performance: Models, Strategy
and Performance Metrics}, Korajczyk, R.A. (editor),  London: Risk
Books, pp. 15-22.



\end{thebibliography}

\end{document}